\documentclass{appolb}
\usepackage{epsfig}

\begin{document}
\title{
Numerical renormalization group study of two-level Kondo effect: 
Discovery of new fixed point
\thanks{Presented at SCES'02}%
}
\author{
M. Kojima, S. Yotsuhashi, K. Miyake
\address{
Department of Physical Science, Graduate School of Engineering 
Science \\
Osaka University, Toyonaka, Osaka 560-8531, Japan
}}
\maketitle
\begin{abstract}
The model of two-level Kondo effect is studied by the Wilson numerical 
renormalization group method.  It is shown that there exist a new type 
of weak-coupling fixed point other than the strong-coupling 
fixed point found by Vladar and Zawadowski two decades ago by means of 
the one-loop and two-loop renormalization group methods.  
\end{abstract}

  
\section{Introduction}
The two-level Kondo (TLK) effect was first studied by Kondo more than two 
decades ago by means of a perturbation calculation~\cite{Kondo}.  
In early 80's, Vladar-Zawadowski (VZ) 
showed on the basis of the poorman's scaling~\cite{VZ1} and two-loop 
renormalization group~\cite{VZ2} methods that the fixed point of the 
TLK model is given by that of the pseudo-spin-1/2 magnetic Kondo model.  
This problem attracted renewed interests in mid 80's in relation to 
the two-channel Kondo (TCK) effect which is characterized by the non-Fermi 
liquid fixed point~\cite{NB,Cox,CZ}.  It is because the Hamiltonian at 
strong-coupling fixed point of TLK system is inevitably mapped to the 
TCK Hamiltonian if the spin-degrees of freedom are read as the channel-degrees 
of freedom.  The purpose of the present contribution 
is to investigate the nature of the fixed point of TLK system itself 
on a more solid calcualtion using the Wilson numerical renormalization group 
(NRG) method.

\section{Model}
We use the same model Hamitonian as that adopted by VZ~\cite{VZ1}:
\begin{equation}
H=\sum_{{\vec k}}\varepsilon_{k}a^{\dagger}_{{\vec k}}
a_{{\vec k}}+\sum_{i=x,y,z}\Delta^{i}\tau^{i}+
2\sum_{{\vec k}_1,{\vec k}_2}\sum_{i=x,y,z}
(a^{\dagger}_{{\vec k}_2}V_{{{\hat k}}_2,{{\hat k}_1}}^{i}
a_{{\vec k}_1})\tau^{i},
\label{eq:1}
\end{equation}
where $a_{{\vec k}}$ denotes annhilation operator of the conduction electron 
with the wavevector ${\vec k}$, $\tau^{i}$ the $i$-th component of the 
pseudo spin describing the two-level (TL) degrees of freedom, 
and the coupling $V$'s are given as follows:
\begin{eqnarray}
V_{{{\hat k}}_2,{{\hat k}_1}}^{z}&=&{{\rm i}V_{z}\over 4\pi}
(a_{0}+3a_{1}\cos\theta_{{\hat k}_{2}{\hat k}_{1}})
({\hat k}_1^{z}-{\hat k}_2^{z}),
\label{eq:2}\\
V_{{{\hat k}}_2,{{\hat k}_1}}^{x}&=&{V_{x}\over 4\pi}
(a_{0}+3a_{1}\cos\theta_{{\hat k}_{2}{\hat k}_{1}})
({\hat k}_1^{z}-{\hat k}_2^{z})^{2}, 
\label{eq:3}
\end{eqnarray}
and $V_{{{\hat k}}_2,{{\hat k}_1}}^{y}=0$, with 
$V_{z}\equiv k_{\rm F}d/2$ and 
$V_{x}=(k_{\rm F}d)^{2}\Delta_{0}\lambda/V_{\rm B}$.  
Otherwise, the notations used in (\ref{eq:2}) and (\ref{eq:3}) are the 
same in Ref.~\cite{VZ1}.  
We restrict the partial wave describing the conduction 
electrons within $\ell=0,1,2,3$ and $m=0$, $m$ being the $z$-component 
of the angular momentum $\ell$, around the TL system as in Ref.~\cite{VZ1}.  

\section{Results}
\begin{figure}[!ht]
\begin{center}
\includegraphics[width=0.85\textwidth]{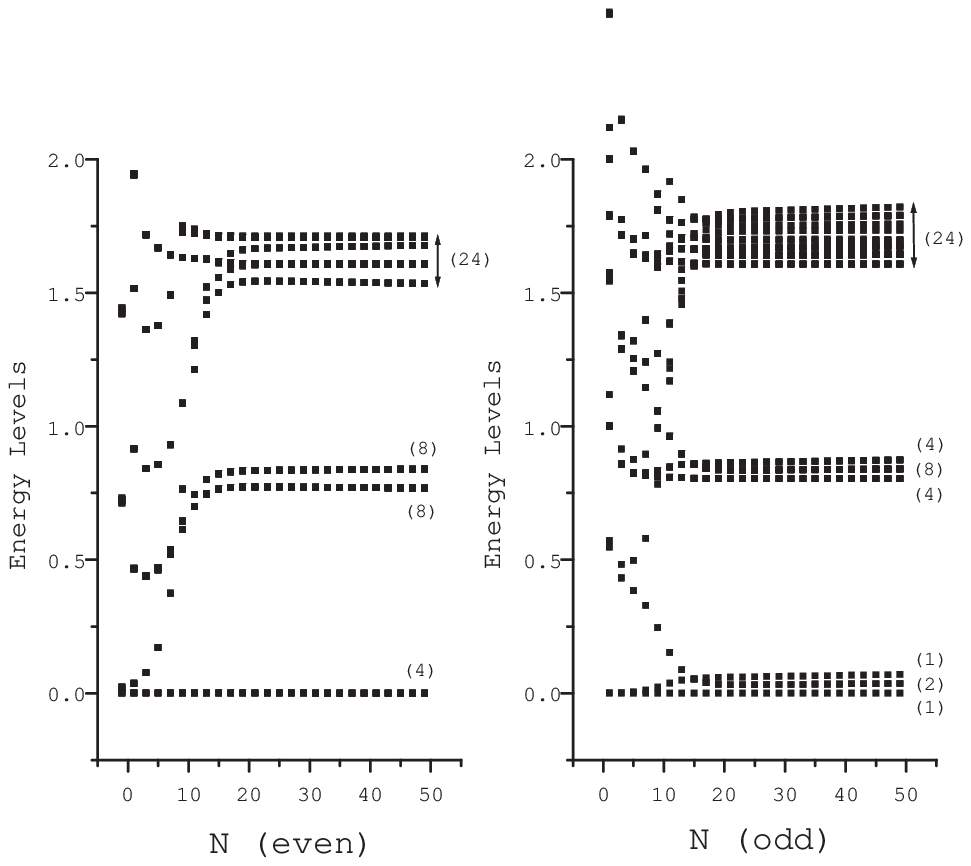}
\end{center}
\caption{Energy flow diagram for the parameter set, $V_{z}=1$, $V_{x}=0.01$, 
$a_{0}=0.6$, and $a_{1}=1.2$. The number in parentheses is the degeneracy. }
\label{fig:1}
\end{figure}

\begin{figure}[!ht]
\begin{center}
\includegraphics[width=0.85\textwidth]{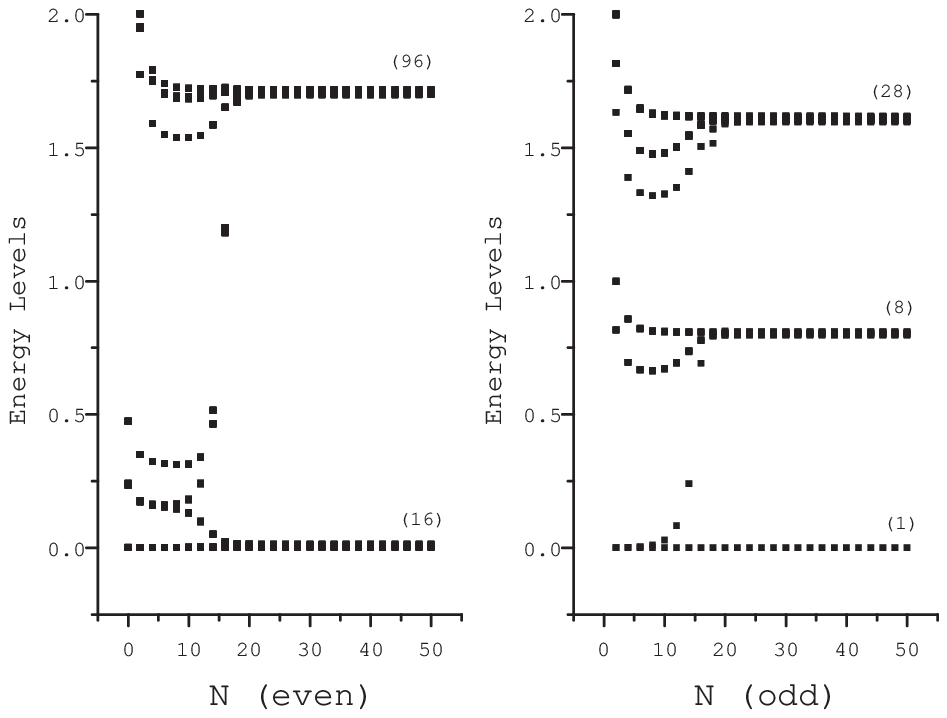}
\end{center}
\caption{Energy flow diagram for the parameter set, $V_{z}=1$, $V_{x}=0.01$, 
$a_{0}=0.6$, and $a_{1}=0.4$.  The number in parentheses is the degeneracy.}
\label{fig:2}
\end{figure}

Energy flow diagrams are classified into two types as shown in Figs.\ 
\ref{fig:1} and \ref{fig:2}.  Fig.\ \ref{fig:1} is for the parameter set, 
$V_{z}=1$, $V_{x}=0.01$, $a_{0}=0.6$, and $a_{1}=1.2$, and corresponds to the 
strong-coupling fixed point although 
it shows no even-odd alternation just as in the TCK effect.  
This is because the conduction electrons have 4-channel 
corresponding to partial wave $\ell=0,1,2,3$, two of which are trapped 
by the TL system.  The excitation energies and 
the degeneracy of each energy levels (shown in parentheses) can 
be explained completely by this picture.  
Small splitting of the energy levels is caused by the growth of the potential 
scattering term.  Such a phenomenon occurs also in the sd-model if the 
potential scattering term is added to the model.  
The scaling unit $\Lambda$ is set as $\Lambda=3$ throughout this paper.  
The number of retained states at each step is 400.  
Fig.\ \ref{fig:2} is for 
the parameter set, $V_{z}=1$, $V_{x}=0.01$, 
$a_{0}=0.6$, and $a_{1}=0.4$, and represents the weak-coupling fixed point 
because it exhibits the even-odd alternation and 
can be understood as the coupling between conduction electrons and 
TL system is absent.  Indeed, the degeneracy (shown in parentheses) can 
be completely explained by this weak-coupling assumption.  

We have searched fixed points for a wide range of parameter sets, 
and determined the phase boundary between two regimes of weak- and 
strong-coupling fixed point as shown in Fig.\ 3.  Although we have 
set $\Delta^{i}=0$ here, the effect of finite field $\Delta^{i}$ does not 
change the phase diagram qualitatively unless $|\Delta^{x}|>|V_{x}|$.  
The weak-coupling region is located along the crossover regime of the two 
types of strong-coupling fixed point of the poorman's scaling equations given 
originally by VZ~\cite{VZ1}.  One type of the strong-coupling fixed points 
has already been found by VZ on the basis of a physical intuition, while 
another type of strong coupling fixed point is found for the frst time in 
this study by solving numerically the coupled 
scaling equations of 48 components: 
\begin{equation}
{\partial V^{s}_{\alpha\beta}\over \partial x}=
-2{\rm i}\sum_{i,j}\epsilon^{ijs}\sum_{\gamma}V^{i}_{\alpha\gamma}(x)
V^{j}_{\gamma\beta}(x),
\label{eq:4}
\end{equation}
where $V^{i}_{\alpha\beta}$'s are the matrix elements of 
$V^{i}_{{\hat k}_{2}{\hat k}_{1}}$, (\ref{eq:2}) and (\ref{eq:3}), 
in the space of $\ell=0,1,2,3$ and $m=0$, and  the scaling variable 
$x\equiv \ln(D_{0}/D)$~\cite{VZ1}.  
Both of the strong-coupling 
fixed points are characterized by the pseudo-spin-1/2 Kondo effect.  
It is noted that the region of weak-coupling fixed point is found only 
through the NRG analysis.  Implication of new fixed point will be discussed 
elsewhere.

\begin{figure}[!ht]
\begin{center}
\includegraphics[width=0.45\textwidth]{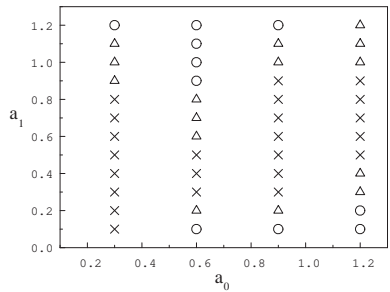}
\end{center}
\caption{Distribution of strong-coupling (circle) and weak-coupling 
(cross) fixed point.  The points presented by triangle cannot be 
identified clearly with each fixed point due to numerical uncertainty.  
}
\label{fig:3}
\end{figure}

This work was supported by the Grant-in-Aid for COE Research (10CE2004) from 
Monbukagaku-sho.

\end{document}